\documentclass[conference]{IEEEtran}

\ifCLASSINFOpdf
   \usepackage[pdftex]{graphicx}
\else
\fi
\usepackage{paralist}
\begin{document}

\title{Do Bots Modify the Workflow of GitHub Teams?}

% author names and affiliations
% use a multiple column layout for up to three different
% affiliations
\author{\IEEEauthorblockN{Samaneh Saadat}
\IEEEauthorblockA{Department of Computer Science\\
University of Central Florida\\
Orlando, Florida\\
Email: s.saadat@knights.ucf.edu}
\and
\IEEEauthorblockN{Natalia Colmenares}
\IEEEauthorblockA{Department of Computer Science\\
University of Central Florida\\
Orlando, Florida\\
Email: natcolchu@knights.ucf.edu}
\and
\IEEEauthorblockN{Gita Sukthankar}
\IEEEauthorblockA{Department of Computer Science\\
University of Central Florida\\
Orlando, Florida\\
Email: gitars@eecs.ucf.edu}}

% make the title area
\maketitle

% As a general rule, do not put math, special symbols or citations
% in the abstract
\begin{abstract}
The ever-increasing complexity of modern software engineering projects makes the usage of automated assistants imperative. Bots can be used to complete repetitive tasks during development and testing, as well as promoting communication between team members through issue reporting and documentation.  Although the ultimate aim of these automated assistants is to speed taskwork completion, their inclusion into GitHub repositories may affect teamwork as well.  This paper studies the question of how bots modify the team workflow. We examined the event sequences of repositories with bots and without bots using a contrast motif discovery method to detect subsequences that are more prevalent in one set of event sequences vs. the other.  Our study reveals that teams with bots are more likely to intersperse comments throughout their coding activities, while not actually being more prolific commenters.

%The ever-increasing complexity of modern software engineering makes the automation of tasks imperative. Bots, that are used by project maintainers to enhance their workflow, may impact team dynamics. We aim to understand whether the use of bots in GitHub teams changes the dynamics of software engineering teams. Event sequences of teams with bots and without bots are examined to reveal differences in team dynamics. We leverage a contrast motif discovery method to detect subsequences (i.e. motifs) that are more prevalent in one set of event sequences vs. the other. 
\end{abstract}

% no keywords

% For peer review papers, you can put extra information on the cover
% page as needed:
% \ifCLASSOPTIONpeerreview
% \begin{center} \bfseries EDICS Category: 3-BBND \end{center}
% \fi
%
% For peerreview papers, this IEEEtran command inserts a page break and
% creates the second title. It will be ignored for other modes.
\IEEEpeerreviewmaketitle

\section{Introduction}
% BOTS
Research on improving the productivity and workflow of software engineering teams has gained traction in recent years ~\cite{meyer2014software,saadat2020analyzing}.
Providing better and smarter tools to help developers simplify and automate tasks is one approach to satisfy that need. A bot is an interface that connects users and services \cite{lebeuf2018taxonomy};  many of them are simple, but more advanced bots use  artificial intelligence techniques. Erlenhov et al.  describe an ideal software bot as "an artificial software developer which is autonomous, adaptive, and has technical as well as social competence" \cite{erlenhov2019current}.
Bots support different software development tasks:  1) automating routine tasks such as merging changes across different branches 2) scheduling activities according to developer preferences,
 3) performing repetitive tasks such as answering user questions, and 4) improving decision making by collecting, analyzing and sharing relevant data.

% BOTS IN TEAMS
In addition to promoting individual productivity, bots can also be used to support team cognition. For example, team situational awareness can be enhanced through the deployment of bots~\cite{poppendieck2003lean}.  Software bots enhance team communication by initiating necessary conversations or reducing the amount of communication by automating tasks \cite{storey2016disrupting}.
Although bots are generally created with the purpose of enhancing  productivity, they may have a negative impact on user experiences \cite{liu2020understanding}. Thus, it is imperative to study the modifications that bots make to the software engineering workflow \cite{wessel2020effects,mirhosseini2017can}. 
While many GitHub bot studies focus on project outcomes \cite{wessel2018power,wessel2020effects}, only a few have considered the developer experience~\cite{liu2020understanding}. 
 In our research, we intend to shed light on the effect of bots on the workflow of human team members in human-bot teams.

% HUMAN-BOT INTERACTION
% Unlike prior studies that focus on a small portion of GitHub repositories \cite{wessel2018power}, we analyzed thousands of repositories.
Many human-bot interaction studies use surveys as their main source of data \cite{liu2020understanding,wessel2018power}. While surveys provide valuable information about user experience, it is challenging to survey thousands of developers. Thus, it is valuable to supplement survey studies with data extracted from mining software repositories. In this paper, with the use of archival data of GitHub activities, we studied thousands of GitHub repositories both with and without bots.

Rather than manually identifying bots in a small set of repositories or limiting our investigation to well known bots, our study relies on automatic bot detection.  Golzadeh et al. used repetitive comments to detect bots in one specific domain (e.g. pull request) \cite{golzadehbot}.  However, there are many bots that do not make any comments. 
To address this issue, we implemented a more comprehensive approach to automatically detect GitHub bots in a large scale.  First, we manually labeled hundreds of accounts to serve as training data for our bot detection classifier.
Our classification approach leverages many features of  an account, in addition to the comment similarity, to identify bots.  

% SEQUENCE
To study team behavior, we apply a sequence mining approach to extract temporal ordering information. 
We created event sequences of GitHub teams with bots (aka human-bot teams) and teams without bots (aka human-only teams). Then we applied our new contrast motif discovery technique (described more fully in \cite{saadat2020contrast}) on this data to discover subsequences that occur more frequently in human-bot vs. human only teams. Note that if the sequence groups are very similar, no contrast motifs will be discovered since they explicitly capture \textit{differences} in datasets rather than just \textit{frequent occurrences}. Our aim is to answer the following research questions about team workflow:

\begin{compactitem}
\item \textbf{RQ1}: do  contrast motifs exist that distinguish human-bot teams from human only ones?
\item \textbf{RQ2}: how do bots affect the team contrast motifs?
\end{compactitem}

\section{Related Work}
\subsection{Bots}
% Software bots
Storey et al. provided a framework to examine the impact of bots on software development teams by categorizing bot roles and their effectiveness in helping developers meet their goals \cite{storey2016disrupting}. The authors urged the research community to study the impact of bots, positive or negative. Since then, research on the impact of automation has grown significantly. A study on a sample of popular repositories before and after bot adoption found no significant difference in metrics such as time to pull request before and after bot adoption \cite{wessel2018power}. 
However, a more recent paper found that the adoption of code review bots increases the number of merged pull request and decreases communication among developers \cite{wessel2020effects}.

% GitHub bots
% Developers neglect to update legacy software dependencies, resulting in buggy and insecure software. One explana- tion for this neglect is the difficulty of constantly checking for the availability of new software updates, verifying their safety, and addressing any migration efforts needed when upgrading a dependency. Emerging tools attempt to address this problem by introducing automated pull requests and project badges to inform the developer of stale dependencies.

In addition to considering the effectiveness of bots towards improving taskwork, some authors have examined social interactions between bots and humans.  Liu et al. found that the \textit{Stale} bot, which helps maintainers triage abandoned issues and pull requests by marking them after a period of inactivity, can create a negative experience for contributors \cite{wessel2019should,liu2020understanding}.   This illustrates the importance of research on human-bot interaction towards providing insights on best practices for the development of effective software bots \cite{storey2016disrupting,brown2019sorry,wessel2018power}.  Desiderata for automation include improving social interactions, better management of the workflow, and increasing the awareness of the developers about bot capabilities. 
%% remove if space is limited
%Erlenhov et al. goes one step further and describe an ideal software bot as "an artificial software developer which is autonomous, adaptive, and has technical as well as social competence" \cite{erlenhov2019current}.

Dey et al.~\cite{dey2020} created a bot classifier, BIMAN, for detecting bots in the World of Code~\cite{worldofcode} dataset; their technique focuses on commit activity.  Here we propose a simpler classifier that is well suited for event data extracted from GHTorrent~\cite{gousios2012}; unlike BIMAN, it does not require tracking file modifications.

%This illustrates the importance of understanding the impact of bots on individuals as well as team processes.

\subsection{Team Sequences}
% SEQUENCES
%Sequences are prevalent in several fields including physical, biological, and computer sciences. Hence, numerous data mining methods have been developed to analyze various types of sequences.
% remove if space is limited
% For example, in molecular biology, the sequencing of nucleotides in a strand of DNA provides a massive amount of sequence data that has motivated scientists to develop methods to examine various aspects of the sequences including sequence assembly, motif finding, etc. %evolutionary? %CITE

% TEAM SEQUENCES
%Team researchers have recognized the dynamic that teams and their tasks have.
Analyzing the temporal sequences of team activities can provide new insights about interactions between team members.  Herndon and Lewis proposed that sequence mining methods are well suited to answer research questions concerning the dynamic nature of teamwork and teams~\cite{herndon2015applying}.
The advantage of sequence-based methods is that they do not isolate a single event, but instead examine the events ``in their continuity'' \cite{herndon2015applying}.

 Process mining approaches~\cite{poncin2011,gupta2014} have been applied to software repositories to capture business workflows.  
 Bingham et al. exploited sequence mining methods to explain differences in the performance of organizations by examining sequence patterns of learning processes \cite{bingham2012learning}.
%bingham2012learning: data show how different learning sequences differentially affect both shorter- and longer-term performance, suggesting that it matters which learning processes are used and when. 
By analyzing sequential changes in work processes of an organization over a period of time, Pentland et al. discovered that variability was negatively related to performance \cite{pentland2003sequential}. 

% A number theories on team development and activity discuss sequence-based ideas and sequence mining methods can be developed to examine these theories. For instance, there is a body of work on group development that convey progress of teams is through a sequence of stages and patterns of these sequences is related to operation and performance of the team in the future \cite{tuckman1965developmental,wheelan1994group}. 

%Herndon et al. state that sequence mining approaches have myriad applications in team research \cite{herndon2015applying}. 
%The authors argue that optimal matching approaches can be used to compare team process sequences for the purpose of examining the impact of weak or strong ties on team processes.
%Optimal matching algorithms measure the distance between two sequences by calculating the cost of transforming one sequence into the other.
%Although optimal matching techniques have widespread use in sociology, their main drawback is that the cost function, which is defined by the researchers, can influence the outcome \cite{robette2012harpoon}.

% remove if space is limited

Certain teamwork studies have failed due to their inability to detect differences in team performance caused by the experimental manipulation.
For instance, Lewis et al.~\cite{lewis2005transactive}, studied the effect of Transactive Memory Systems (TMS) on team performance. They hypothesized that enabling TMS in teams would translate into higher performance. Although they observed higher performance for teams with TMS vs. without, they also observed that the highest performance occurred in teams with TMS disabled first and then enabled.
\cite{herndon2015applying} argues that the difference in the performance of these teams is probably caused by dynamic processes and suggests that sequence mining methods could be used to further analyze the differences.

These studies manually coded states of teams and their activities during a predefined time interval (e.g. 1 minute).
However, with GitHub, a large corpus of team activities is  effortlessly accessible. Our work builds on a study of GitHub repository event sequences that revealed that teams that use bots have less repetitive sequences of events \cite{saadat2020explaining}.  In this paper, we use contrast motif discovery to identify the specific workflow differences caused by the introduction of bots.

%%%%%%%%%%%%%%%%%% herndon2015applying
% [Samaneh:] Summary
% 1) Using optimal matching to find how similar sequences of teams are
% 2) clustering sequences to reveal hidden patterns

% [Samaneh:] In GitHub, We don't have any information about the duration of each event. but number of times an event is repeated in a period could be sth to consider.
% [Samaneh:] parameters of our algorithm does not significantly change the outcome.
%%%%%%%%%%%%%%%%  herndon2015applying

\section{Bot Detection}
\subsection{Data}
We obtained the most recent data month available from the GHTorrent~\cite{gousios2012} dataset (June 2019). GHTorrent~\cite{gousios2012} is an offline mirror of GitHub public event data. 

\subsection{Labeling}
In the dataset, we found $4815$ accounts with "bot" in their username.
We randomly selected $612$ of these accounts for manual labelling.   To ascertain if the account belonged to a bot, we conducted a web search and also read the GitHub profile. 
Out of these $612$ GitHub accounts, we verified that $424$ of the accounts (about $70\%$) belonged to bots. 
We have made our GitHub dataset and manually labeled accounts publicly available\footnote{https://osf.io/de6y7/}

\subsection{Features}
The following features were extracted for all the accounts in our dataset to be used as the input of our bot detection classifier:
\begin{itemize}
    \item \textbf{Comment similarity} is the average cosine similarity of comments.  This feature is relevant because bots tend to follow a template for their comments so comment similarity is high among bot accounts. Cosine similarity of $-1$ denotes that no comments were found to compare.
\item \textbf{Organization owned} shows if the account has ever belonged to an organization. Our investigation of the data showed that accounts that have "bot" in their name have a higher probability of being a bot if they are owned by an organization.
\item \textbf{Unique event types} denotes the category  of GitHub events the account generated (e.g. if an account only performs \textit{issue comments} events this feature has a value of 1). Bot accounts tend to perform a limited number of event types while human software developers inevitably generate a diverse distribution of events. 
\item \textbf{Placement of "bot"} is the location of the string “bot” in the account name: beginning, middle, or end.
\end{itemize}

\subsection{Bot Detection Classifier}
We trained and tested various classifiers on the labeled dataset including \textit{Logistic Regression}, \textit{Random Forest}, and \textit{Gradient Boosting}.
Classifiers were evaluated using cross validation to ensure the model is not overfitted. 
Since the dataset is imbalanced, we used a stratified $k$-fold cross validator to preserve the percentage of samples for each class.
$k$ was set to $5$ in our experiments.

All tested classifiers achieved a F1-score higher than 0.9.
Gradient Boosting achieved the highest F1-score of 0.93 (precision=0.89, recall=0.97).
We used Gradient Boosting to learn a model on the hand labeled 612 samples which we then used to label the remaining accounts.

\section{Team Event Sequence Analysis}
\begin{figure*}
\includegraphics[width=0.7\textwidth]{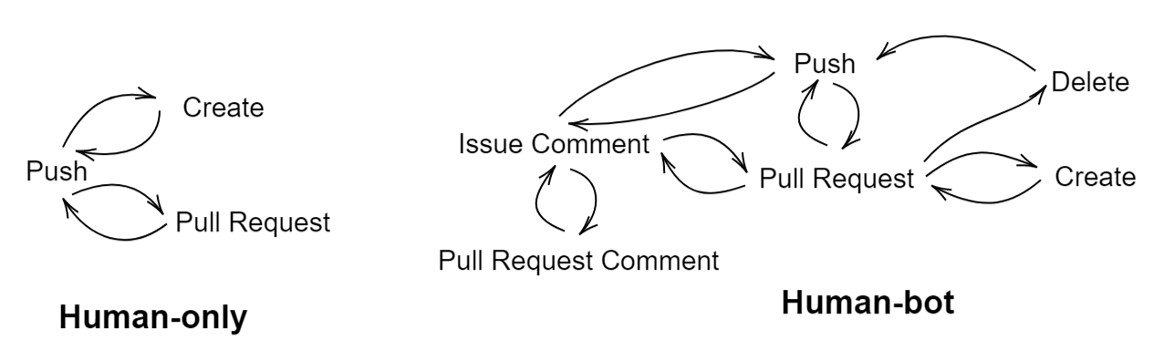}
\centering
\caption{ 
Motif graph for human-only and human-bot teams.  Teams with bots are more likely to
intersperse comments throughout their coding activities, while
not actually being more prolific commenters.
} 
\label{fig:gh_motif_graph}
\end{figure*}

\subsection{Event Dataset}
More than 46 million events were found during the month of June 2019. % exact events count = 46,260,306
These events were executed by more than 420,000 active users over 285,000 repositories. We denote a user or repository as active if they perform at least one push or pull request.
A GitHub user is considered a member of a team if they contributed at least one push or one pull request event to the repository \cite{saadat2020analyzing}.
Events related to bots and non team members were removed from the dataset. A repository with at least two members is designated as a team. We separated the data into two types: (1) human-bot teams using automated accounts and (2) human-only teams that are not using any automated services for their repository.

\subsection{Team Sequences}
We ordered the events corresponding to each team based on the time that event is performed and created event sequences for each team. 
Table~\ref{tab:event_prop} shows the proportion of different event types in our dataset. \textit{Push} and \textit{pull request} are the most frequent events. These two events are used to propagate changes made on a local repository to the main repository.

\begin{table}
\centering
\caption{The percent of different event types in our dataset.}
\begin{tabular}{ll}
\hline
\textbf{Event Type }                   & \textbf{Percent} \\
\hline
Push                     & 44.1\% \\
Pull Request             & 15.6\% \\
Create                   & 11.3\% \\
Issue Comment             & 9.1\% \\
Pull Request Review Comment & 5.6\% \\
Delete                   & 4.7\% \\
Issues                   & 4.0\% \\
Watch                    & 2.4\% \\
Fork                     & 1.3\% \\
Release                  & 0.56\% \\
Gollum                   & 0.38\% \\
Member                   & 0.31\% \\
Commit Comment            & 0.16\% \\
Public                   & 0.07\%\\
\hline
\end{tabular}
\label{tab:event_prop}
\end{table}

% Cleaning
Since events with low frequencies will not appear in the discovered motifs, we removed the least frequent events to make the algorithm run faster. 
Additionally, we combined \textit{issue event} and \textit{issue comment} as they both refer to the same kind of activity. 
Our final list contained six event types including: \textit{push}, \textit{pull request}, \textit{issue}, \textit{pull request review comment}, \textit{create}, and \textit{delete}.

After transforming sequences, we removed sequences shorter than five to guarantee that the sequence length is longer than the window length and tested window sizes longer than two.
As the threshold goes higher, more and more teams are omitted from the data set.
Five was selected as the threshold as it removes very short sequences without discarding a large number of teams.

\subsection{Team Sampling}
Our dataset is imbalanced with 201,204 human-only teams and 4,205 human-bot teams. We down-sampled human-only teams to make a fair comparison between the two sets of teams \cite{saadat2020explaining}.
For each team, we create a vector that contains the frequency of the different event types in that team.
For each human-bot team, we found the most similar human-only team with respect to their event frequency vector and down-sampled the human teams to 4,205 teams corresponding to the 4,205 human-bot teams.

Table \ref{tbl:event_medians} shows the median event frequencies. 
The first column corresponds to human-bot teams. 
The second column contains the medians of human-only teams events before down-sampling.
The median values in these two columns are relatively different.
The third column shows medians after the down-sampling of human-only teams; these values are close to human-bot teams.
This demonstrates the effectiveness of our sampling approach at creating sets of sequences with comparable unigram frequencies.  Hence any discovered differences result from event ordering, rather distribution differences.

\begin{table}
\centering
\caption{Median value of event frequencies before and after sampling.}
\begin{tabular}{l|lll}
\hline
\textbf{Event Type}   &  \textbf{Human-bot}	&\textbf{Human} &	\textbf{Down-sampled Human}\\
\hline
\textbf{Push}                     & 11.0 & 9.0 & 12.0 \\
\textbf{Pull Request}              & 9.0  & 4.0 & 9.0  \\
\textbf{Issues}                   & 1.0  & 0.0 & 1.0  \\
\textbf{Issue Comment}             & 6.0  & 0.0 & 5.0  \\
\textbf{Create}                   & 2.0  & 3.0 & 2.0  \\
\textbf{Delete}                   & 1.0  & 1.0 & 1.0  \\
\hline
\end{tabular}
\label{tbl:event_medians}
\end{table}

\subsection{Contrast Motifs}
To detect differences between sequences of human-bot and human-only teams we utilized a \textit{Contrast Motif Discovery} technique developed in our prior work~\cite{saadat2020contrast}. 
This approach discovers short subsequences (aka motifs) that are significantly more similar to one set of sequences vs. other sets.  
The algorithm first finds a set of motifs for each group of sequences and then refines them to include the motifs that distinguish that group from other groups of sequences. Motifs are only selected 1) if the  average distance to the sequences in their group is lower than the distances to the sequences from other groups and 2) this difference is statistically significant.  

We ran the \textit{Contrast Motif Discovery} tool on the sequences we created for GitHub teams.
We tested window sizes from $2$ to $5$. As the window size becomes larger, repetitive patterns emerge within motifs. Window size $4$ was the largest window size with the least repetition. Therefore, in this study, we present our results for a window size of $4$.

% FINDINGS

Figure \ref{fig:gh_motif_graph} shows the graph representation of the contrast motifs for each team type. We observe a more complex graph structure for human-bot teams while human-only teams have a simpler structure. This observation indicates that the repetitive patterns in human-bot teams are more complex than human-only teams. Note that this does not indicate that these patterns do not occur in human-only teams, but that they are significantly more frequent in human-bot teams.

Another observation about the graphs shown in Figure \ref{fig:gh_motif_graph} is that in human-bot teams, issue comments occur before and after each event type. This does not mean that human-bot teams make more issue comments since we selected repositories in a way to maintain similar event frequencies. We hypothesize that this pattern occurs because in human-bot teams issue comments are intermixed with  other events rather than being clustered together. % what are the benefits of this for teams?
To confirm this hypothesis, we examined the length of consecutive issue comment events in human-only and human-bot teams. 
First, we measured the average length of consecutive issue comment events in all sequences. Then, we ran a \textit{Mann-Whitney U-test} analysis to investigate whether there is a difference between two types of teams.
Our \textit{U-test} indicates that the average length of consecutive issue comment events in human-only teams is higher than this value in human-bot teams and the difference is statistically significant (\textit{p-value}$=10^{-7}$).
This finding means that bots force human members of the team to discuss issues between different stages of the workflow.

% Issues in GitHub repositories can be used to develop an understanding of the problem at hand and the work needed to resolve the problem, and also offer a space for the discussion and evaluation of potential solutions and strategies to address needs of users and the project.

Literature shows that when automation is introduced, humans lose situation awareness (SA) \cite{endsley1996automation}. When humans serve as monitors of automated systems, humans become slower to detect and fix problems since they are not the actors of the automated tasks.
Our interpretation of commenting patterns in human-bot teams is that human team members may use comments to stay in the loop and as a mechanism to gain situation awareness. 
It is crucial to further study the performance of human-bot teams to ensure that continuous communication compensates for the loss of situation awareness and does not negatively impact the performance of teams.

\subsection{Limitations}
Our methodology has certain limitations which pose threats to validity.  Our bot classifier may fail to detect certain categories of bots; hence the extracted contrast motif may only characterize the interaction patterns of bots detected by our classifier.
Also the matching principle used when sampling human-only teams allows us to control the variables being matched but threatens external validity.

\section{Conclusion and Future Work}
This paper studies the following question: how do bots modify the workflow of software engineering teams?
For this study, we constructed a bot detection classifier to discover GitHub bots and to identify the repositories that use them. 
To compare human-bot teams with human only ones, we applied our contrast motif discovery technique to the two groups of sequences.  \textbf{RQ1} was answered affirmatively; both human-only and human-bot teams possess contrast motifs that are found predominantly in only one of the groups.

Our analysis of the contrast motifs indicates that in human-bot teams,  issue comments are scattered throughout the event sequences while in human-only teams the contrast motifs are simpler and issue comments tend to be clustered together (\textbf{RQ2}). The continuous communication that we observe in human-bot teams could occur due to a lack of situation awareness caused by introduction of automation.
We plan to measure and compare the performance of human-bot and human-only teams to understand the impact of automation and communication differences on the outcome of the teams. We hypothesize that teams with bots are more likely to intersperse comments throughout their coding activities, while not actually being more prolific commenters.   Note that the type of bot obviously affects the workflow.  In this study, we aggregated sequences generated by teams using many types of automation.  It may be beneficial to do a narrower study examining the effects of a single type of bot.

%Another avenue for future work is the use of contrast motifs for simulating GitHub users. 
%Prior studies discovered clustering archetypes of GitHub users and simulated GitHub activities in a future time period based on the average and standard deviation of activities of users in the clusters \cite{saadat2018initializing}.
%We believe finding contrast motifs for each user archetype and having agents to mimic the motifs can enhance the accuracy of the simulation as motifs provide frequent ordering of events. 

Kozlowski et al.'s normative model of team development defines four continuous and overlapping phases: team formation, task compilation, role compilation, and team compilation \cite{kozlowski1999developing}. In future work, we will extract contrast motifs from different phases of team development in order to compare and contrast the processes in these phases across team types. 
%This analysis can help us to understand whether there are phases when human-bot and human-only teams behave similarly; we could also identify when their behavior begins to diverge.

\section{Data Availability}
Our dataset is publicly available at: https://osf.io/de6y7/

\section*{Acknowledgment}
This work was supported by grant W911NF-20-1-0008 from DARPA. The views and opinions contained in this article are the authors and should not be construed as official or as reflecting the views of the University of Central Florida, DARPA, or the U.S. Department of Defense.

\bibliography{main}
 \bibliographystyle{plain}

% that's all folks
\end{document}